\documentclass[%
 reprint,
 amsmath,amssymb,
 aps,onecolumn
]{revtex4-2}

\usepackage{graphicx} 
\usepackage{dcolumn} 
\usepackage{bm} 
\usepackage{comment}
\usepackage{cancel}

\usepackage{xcolor}

\begin{document}

\title{State Transfer in Noisy Modular Quantum Networks}

\author{Markku Hahto}
\affiliation{%
 Department of Physics and Astronomy, University of Turku, FI-20014, Turun Yliopisto, Finland.
}%
 
\author{Jyrki Piilo}%
\affiliation{%
 Department of Physics and Astronomy, University of Turku, FI-20014, Turun Yliopisto, Finland.
}%

\author{Johannes Nokkala}%
\affiliation{%
 Department of Physics and Astronomy, University of Turku, FI-20014, Turun Yliopisto, Finland.
}%

\date{July 2, 2024}

\begin{abstract}
Quantum state transfer is the act of transferring quantum information from one system in a quantum network to another without physically transporting carriers of quantum information, but instead engineering a Hamiltonian such that the state of the sender is transferred to the receiver through the dynamics of the whole network. A generalization of quantum state transfer called quantum routing concerns simultaneous transfers between multiple pairs in a quantum network, imposing limitations on its structure. In this article we consider transfer of Gaussian states over noisy quantum networks with modular structure, which have been identified as a suitable platform for quantum state routing. We compare two noise models, affecting either the network topology or the network constituents, studying their effects on both the transfer fidelities and the network properties. We find that the two models affect different features of the network allowing for the identification and quantification of the noise. We then use these features as a guide towards different strategies for the compensation of the noise, and examine how the compensation strategies perform. Our results show that in general, modular networks are more robust to noise than monolithic ones.
\end{abstract}

\maketitle


\section{Introduction} \label{sec:introduction}

Quantum state transfer (QST) is an essential part of modern and future quantum technologies \cite{kandel2021adiabatic,laracuente2022modeling,lewis2023low,xiang2024enhanced}. The aim of QST is to send the state of one system to a target system through an intermediary system of similar nature, such as a chain of interacting spins \cite{bose2003quantum}, to avoid shuttling individual systems or the need to convert, e.g., solid state qubits to photonic qubits and vice versa. The intermediary system can be modeled as a network of subsystems and the transfer happens through the Hamiltonian dynamics of the network \cite{bose2003quantum}. In order to achieve unit fidelity---perfect transfer---the network must in general be from a very restricted class of networks \cite{de2005perfect,christandl2005perfect}. Alternatively, transfer may be achieved slowly over random networks by using a weak coupling and taking advantage of a normal mode separated from the rest, which under mild conditions appears by construction \cite{plenio2005high,wojcik2007multiuser,paganelli2013routing,nicacio2016coupled}. QST has been considered in fermionic \cite{bose2003quantum} and bosonic \cite{plenio2005high,chudzicki2010parallel} networks, i.e. spins and harmonic oscillators.
A generalization of QST is quantum state routing \cite{wojcik2007multiuser,paganelli2013routing,yousefjani2020simultaneous,hahto2024transfer}, in which any pair of users can communicate over the network and multiple transfers may happen simultaneously.
Importantly, while perfect QST can be achieved through a single judiciously engineered real Hamiltonian, perfect routing cannot---even if sequential transfers are allowed \cite{kay2011basics}. Allowing complex coupling coefficients limits one to a completely connected network between all users \cite{kay2011basics}. When using normal modes, requiring simultaneous transfers sets new conditions; while a single useful mode suffices for QST, routing requires multiple.
Modularity was identified as the key feature of random networks to provide such modes in \cite{hahto2024transfer}; such networks consist of clearly defined communities with few links between them. More generally, having a community structure is a ubiquitous feature of complex networks \cite{fortunato2010community,fortunato2016community,javed2018community}, providing them distinct features at the mesoscopic level.

In realistic scenarios the network is not an isolated ideal system. It may be subject to various types of noise, and the effect is virtually always detrimental, as seen for imperfections in the network nodes \cite{rafiee2013noise,de2005perfect,coutinho2022robustness,bugalho2023distributing, babukhin2022effect,zwick2014optimized,burgarth2005perfect}, or environment induced decoherence \cite{qin2015high,nicacio2016coupled,kay2016quantum,qin2015protected}. In particular, noise in the links of a spin chain has been studied extensively and various ways for its mitigation have been proposed. These include dynamical modulation applied at the ends which was also found to improve the trade-off between transfer speed and fidelity \cite{zwick2014optimized}, transferring a logical qubit which can also mitigate the harmful effect of timing errors \cite{kay2016quantum}, or using heralded transfer which however makes the transfer time non-deterministic \cite{burgarth2005perfect}.  In hardware implementations unwanted crosstalk between qubits can also be thought of as a special type of link noise, which was found to impact chiefly the time of maximum transfer fidelity, whereas gate errors were found to affect mostly the fidelity maximum \cite{babukhin2022effect}. Environment induced noise has been proposed to be mitigated by increasing the coupling strength of links \cite{qin2015high} or by logical qubits \cite{qin2015protected}. A notable exception is Ref.~\cite{rafiee2013noise} where it was found that adding time dependent noise to the links of a completely connected network can somewhat improve transfer fidelity by essentially partially blocking wrong paths. Neither continuous-variable quantum information nor complex networks have received much attention.

In this article we address this gap by considering transfer of zero mean Gaussian states over networks of quantum harmonic oscillators with a defined community structure, and which are subject to noise. We consider noise in both the network topology and in the network oscillators, and show how they affect transfer over different normal modes of the network. We observe opposite behavior for the two types of noise considered, related to the community structure: while the center-of-mass mode is known to be by construction insensitive to link noise we find that the ability to route may suffer, whereas noise in node frequencies rapidly degrades its ability to support transfer but ability to route degrades gracefully. We show that information at just the mesoscopic level of communities suffices to mitigate the harmful effects of noise. Finally, we apply this feature to develop noise mitigation strategies that avoid the in principle possible but costly reconstruction of full network topology \cite{nokkala2016complex}. Unlike open or closed chains proposed for routing before \cite{wojcik2007multiuser,paganelli2013routing,yousefjani2020simultaneous}, we find that modular networks are robust to losing random links and can maintain a favorable normal mode structure as their size is increased.

The article is structured as follows: in Section~\ref{sec:model} we present the model of the oscillator network and mathematical details relevant for
transfer and routing, as well as the model used to construct the random network structure. In sections~\ref{sec:topologynoise} and \ref{sec:frequencynoise} we show how the different noise models affect the networks properties and discuss some methods of compensating for and identifying the noise. We discuss our results and conclude in Section~\ref{sec:conclusions}.


\section{Model} \label{sec:model}

We consider a network of $n$ quantum harmonic oscillators with unit mass. The position and momentum operators of the oscillators are $\mathbf{q}^\top = \{q_1, q_2,...,q_n\}$ and $\mathbf{p}^\top = \{p_1, p_2,...,p_n\}$ respectively. The oscillators interact with springlike couplings of constant magnitude $g$. The network Hamiltonian is then
\begin{align}
    H = 
        \frac{ \mathbf{p}^\top \mathbf{p} }{2} 
        + \frac{ \mathbf{q}^\top (\omega_0^2 \mathbf{I} + g \mathbf{L}) \mathbf{q} }{2},
\end{align}
where $\omega_0 = 1$ is the bare frequency of the oscillators, and $\mathbf{L}$ is the Laplacian matrix of the network, encoding the information about the network topology. $\mathbf{L}$ is defined by $\mathbf{L} = \mathbf{D} - \mathbf{A}$, where $\mathbf{D}$ is a diagonal matrix containing the degrees of the nodes and $\mathbf{A}$ is the adjacency matrix of the network, whose entries are $A_{ij} = 1$ if oscillators $i$ and $j$ are coupled, and 0 otherwise \cite{wilson2010introduction}.

The network Hamiltonian can be diagonalized into a basis of non-interacting oscillators with an orthogonal matrix $\mathbf{K}$ consisting of the eigenvectors of the Hamiltonian, with which the normal mode position operators are given by $\mathbf{Q} = \mathbf{K}^\top \mathbf{q}$. We denote the normal mode frequencies by $\Omega_i$. In a network with homogeneous oscillator frequencies the center-of-mass mode $\Omega_0$ has the same frequency as the network oscillators, $\Omega_0 = \omega_0$, and the corresponding eigenvector has constant overlap with each network oscillator, $K_{i0} = \frac{1}{\sqrt{n}}$. Notably these are independent of the network topology. This can be seen by considering $\mathbf{L}\mathbf{1}$, with $\mathbf{1}$ being a vector of ones. Each row of the adjacency matrix $\mathbf{A}$ sums to the degree of the node, and since $\mathbf{L} = \mathbf{D} - \mathbf{A}$, each row of $\mathbf{L}$ sums to 0, leading to $\mathbf{L}\mathbf{1} = 0\cdot\mathbf{1}$. Thus $\mathbf{1}$ is an eigenvector of $\mathbf{L}$ with eigenvalue 0, and for the Hamiltonian we have $(\omega_0^2\mathbf{I} + \mathbf{L})\mathbf{1} = \omega_0^2\mathbf{1} + 0\cdot\mathbf{1}$. By normalization the eigenvector corresponding to the eigenvalue $\omega_0^2 \equiv \Omega_0^2$ is then $\frac{1}{\sqrt{n}}\mathbf{1}$.

To this network we add with linear couplings two additional quantum harmonic oscillators, the sender $S$ and the receiver $R$. These contribute to the Hamiltonian the interaction terms $H_I = H_{I,S} + H_{I,R} = -k_S q_S q_i -  k_R q_R q_j$, where $k_{S,R}$ are the coupling strengths of the sender and receiver, $q_{S,R}$ their respective position operators, and $q_{i,j}$ the position operators of the network nodes the sender and receiver couple to. In the normal mode basis this is then written as $H_{I,S} = -k_S q_S \sum_{j=0}^{n-1} \mathbf{K}_{ij} Q_j$, and similarly for the receiver. To have an effective coupling strength of $g$ to normal mode $j$, we must tune $k$ so that $k/K_{ij} = g$. 

We mainly study the transfer of a squeezed vacuum state, $|r=1,\phi=0\rangle$. We also show that the results generalize to other states by considering transfer of entanglement of a two-mode squeezed vacuum state in Appendix~\ref{app:entanglement}. As the Hamiltonians are quadratic the dynamics preserves the Gaussianity of the initial state \cite{ferraro2005gaussian,adesso2014continuous}, which is of the general form $|r,\phi\rangle\otimes\rho_{H}\otimes|0\rangle$ where $\rho_{H}$ and $|0\rangle$ are the initial states for the network and receiver respectively. In simulations $\rho_{H}$ is always taken to be the product of vacuum states of the network oscillators, i.e. $\rho_H = |0\rangle^{\otimes n}$. We additionally remark that it suffices to consider the covariance matrices as the first moments vanish for the chosen states.

In order to transfer a state from the sender to the receiver the coupling strengths and transfer time must be chosen suitably. These were derived for coupled harmonic oscillators in \cite{hahto2024transfer}. The values in the ideal case where the sender and receiver interact with only a single mode are
\begin{align}
    g = \frac{\sqrt{2}}{2c+1}\omega_i^2, \quad t = (2c+1)\pi\omega_i^{-1},
\end{align}
with $c \geq 1$ an integer parameter that sets the transfer time and $\omega_i$ the frequency of the external oscillators, tuned to resonance with a normal mode $\Omega_i$ of the network. A larger $c$ results in a weaker coupling and longer transfer time, which generally improves the transfer fidelity. In this article we have chosen $c = 50$, which numerically works out to $g \approx 0.014 \,\omega_i$. A smaller value could lead to low transfer fidelities over channels which are capable of high fidelity transfer, while a value too large would mean that the receiver has to wait a long time.

Due to deviations from the ideal case---the oscillators interacting with other normal modes in addition to the intended mode, and noise present in the system---the time at which the transfer fidelity reaches its maximum differs slightly from the ideal time, and the variation is different between network realizations. While one possible approach is to search for the maximum fidelity over a large time window $[0,t_{max}]$ as done in \cite{bose2003quantum}, this would be infeasible in practice: the receiver must measure the state before knowing the fidelity, which stops the ongoing transfer. The transfer would then have to be repeated several times to find a time at which the fidelity is maximal. Our approach is to restrict our search to a short time window after the ideal transfer time, $[t_{ideal}, t_{ideal} + \Delta t ]$, which can be combed in a smaller amount of transfers. Additionally as shown in \cite{hahto2024transfer}, while the fidelity at $t_{ideal}$ may be low, the transfer efficiency (e.g. fraction of squeezing received) can still be high due to phase differences between the sent and received state affecting only the fidelity---if the phase difference can be measured and reconciled, one can use the ideal transfer time and match the phase, removing the need for transfer time optimization.

Modular networks are of particular interest to quantum state routing because a network consisting of $N$ distinct communities has $N$ normal modes which are well separated from the rest \cite{chauhan2009spectral}, and with suitable network parameters these normal modes are also separate from each other. The lowest frequency mode $\Omega_0$ by construction has equal overlap with all the network nodes, allowing high fidelity transfer between all pairs of nodes, whereas the coupling strengths of the nodes to the next $N-1$ normal modes are mostly dependent on the community of the node \cite{hahto2024transfer} which restricts the communities between which transfer is possible. As will be seen, while both the slowest and higher frequency modes are robust to link noise, only the latter are robust also to frequency noise. This gives modular networks an edge over monolithic ones not only for routing but also for QST under certain types of noise. 

Here we focus on modular networks generated with the stochastic block model (SBM), which joins random networks with random links to produce a modular network \cite{holland1983stochastic}. SBM takes as parameters the number of communities $N$, the number of nodes in each community $n_c$, and the $N \times N$ probability matrix, which has on the diagonal the probabilities of links within each community $p_{int}$ and on the off-diagonals the probabilities for links between communities $p_{bet}$. We fix the parameters to be $N = 4$, $n_c = 10$, $p_{int} = 0.75$ and $p_{bet} = 0.025$, unless stated otherwise. When calculating features of the network, e.g. normal mode frequencies, we create 1000 realizations of SBM networks with the \textit{NetworkX} Python package \cite{hagberg2008exploring}, again unless otherwise stated. However, when calculating transfer fidelities we limit the number of realizations to 100 since the numerical calculation of the fidelities is more time consuming.

The scope of this article is limited to static noise: the initial network is generated and the corresponding transfer parameters---transfer time and coupling strength---are calculated. The noise is then applied, leading to a network with slightly altered features and thus different eigenvectors and eigenfrequencies. Transfer is attempted over the noisy network using the parameters from the initial network, usually leading to a lower transfer fidelity.

\begin{figure}
    \centering
    \includegraphics[width=0.5\linewidth]{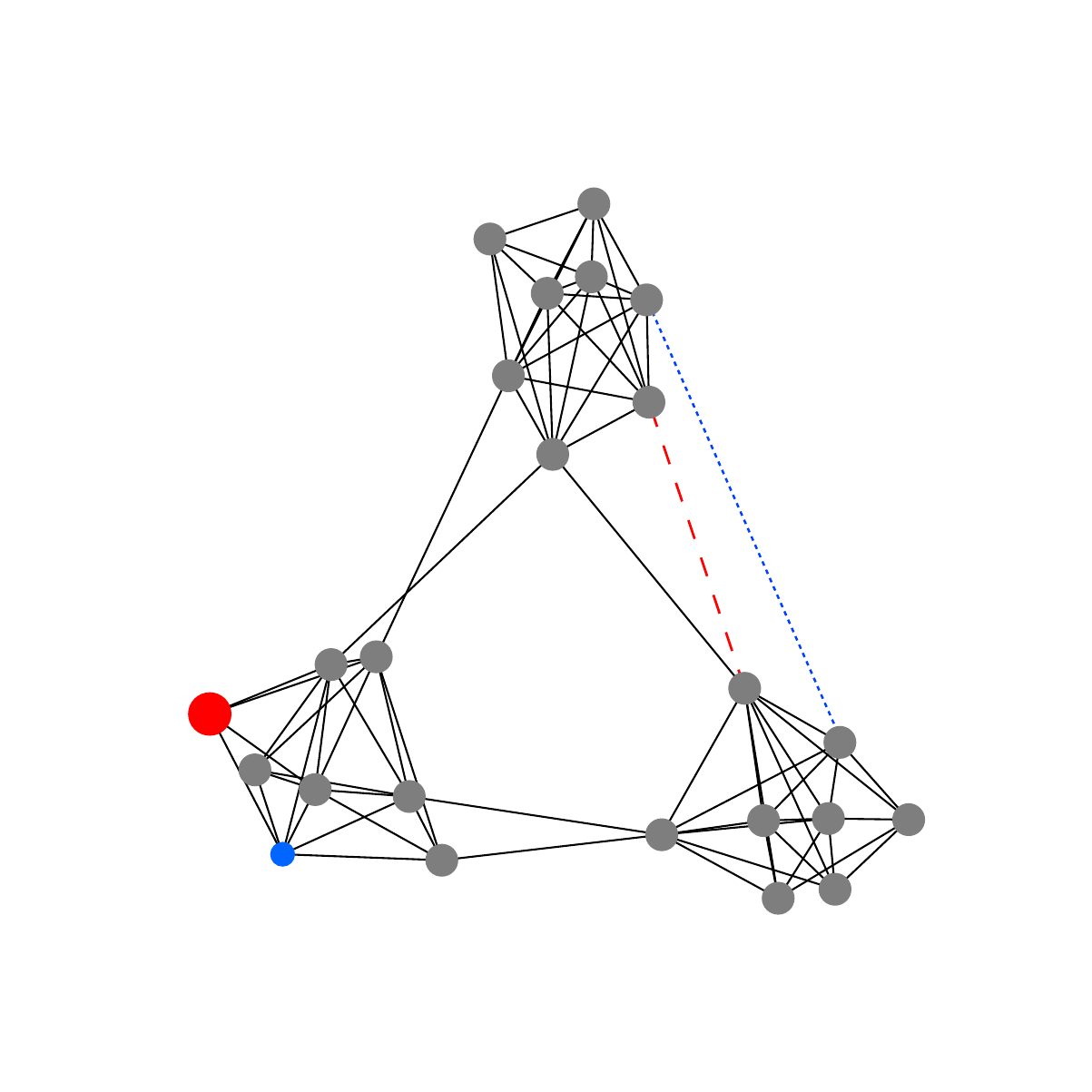}
    \caption{A modular network with three communities of eight nodes each. The dashed red line indicates a link that has been lost, as considered in Section~\ref{sec:topologynoise}, and subsequently compensated for with the dotted blue link. Similarly the large red node represents an oscillator with a detuned frequency, as considered in Section~\ref{sec:frequencynoise}, and its compensation is indicated by the small blue node.
    }
    \label{fig:topo_info}
\end{figure}


\section{Noise in network topology} \label{sec:topologynoise}

In this section we consider noise in the network topology, specifically removal of links from the network as shown in Figure~\ref{fig:topo_info}. Even though deletion of nodes from the graph could be examined, it coincides with the case of removing all the links connecting a single node and considering the rest as the whole network. As we will soon see, this is more extreme than is necessary for interesting features to emerge. Additionally, removing a node alters the eigenvector corresponding to the center-of-mass mode of the Hamiltonian, which we wish to keep untouched for now. We could alternatively consider adding new links in the network, but our testing shows that those results do not qualitatively differ from the case where links are removed.

In a modular network links fall into two distinct categories: links which connect nodes belonging to the same community, and links which bridge two communities. We shall call these \textit{internal links} and \textit{inter-community links} respectively. Changes in the inter-community links have a greater effect on the overall topology of the network, since in a network with pronounced community structure there are plenty of internal links and only a few inter-community links. For a concrete example, the expected fraction of internal links is over $85 \%$ with our default parameter values.

\begin{figure}
    \centering
    \includegraphics[width=0.5\linewidth]{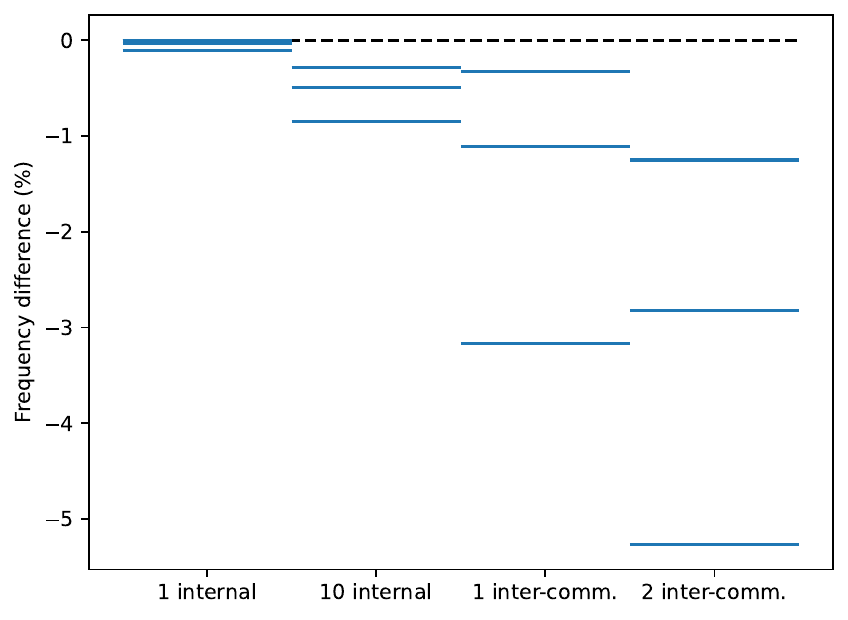}
    \caption{
    Removing links leads to decrease of the frequency of the higher normal modes. Here the frequency difference of modes $\Omega_1-\Omega_3$ are shown for removal of either internal or inter-community links for an ensemble of 1000 SBM networks. The frequency differences for each network realization are sorted and then averaged over the ensemble to show the behaviour of the  most shifted as well as the least shifted frequencies. The y-axis shows the relative frequency difference.
    }
    \label{fig:topofreqs}
\end{figure}

Removing a link decreases the frequencies of the normal modes of the network: for $i > 0$, $\Omega_i' < \Omega_i$ where $\Omega_i'$ are the normal mode frequencies after the removal of a link \cite{porto2017eigenvalue}. Provided the network stays connected, the slowest normal mode $\Omega_0$ is unaffected due to its independence from the network topology. To quantify the difference in importance between internal and inter-community links, in Figure~\ref{fig:topofreqs} we compare the shifts of the frequencies of the slowest $N-1$ affected normal modes when different amounts of links are lost. We see that removing a single inter-community link is more impactful than removing several internal links. In this light, we focus only on the loss of a single link between communities; removing two or more inter-community links would unnecessarily complicate the situation.

When a transfer network has experienced such loss, a natural question is if the losses can be compensated for or if their effects can mitigated without fully reconstructing the network. We show that in the case of a single lost inter-community link, adding any link between the correct communities restores the transfer fidelities to near their original levels. This approach is justified by the fact that all nodes within one community have nearly equal overlaps with any of the $N$ slowest normal modes and thus the effect of connecting any two nodes in two communities has very similar effect on those modes.

\begin{figure}
    \centering
    \includegraphics[width=0.5\linewidth]{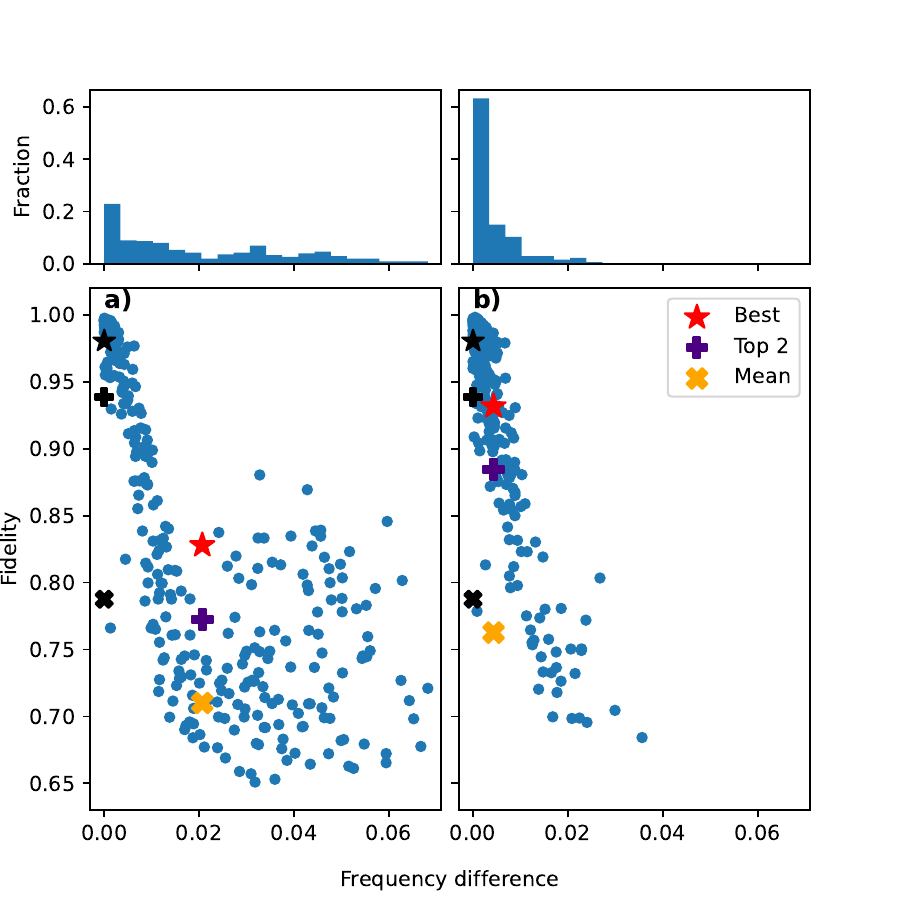}
    \caption{
    Transfer fidelities achieved over the normal modes $\Omega_1-\Omega_3$ compared to the amount the normal mode frequency has shifted. \textbf{a)} A single inter-community link is removed from the network. \textbf{b)} The missing link has been compensated for by adding a different link between the same communities.  The blue markers show the fidelities of the best performing community in each realization and over each of the three normal modes, and the coloured large markers show the averages over the whole ensemble. The black markers show the corresponding fidelities in the lossless case. The histograms over both panels show the distribution of frequency differences.\\
    \textit{Best}: mean fidelity in the community in which the mean transfer fidelity is highest. \textit{Top 2}: mean fidelity in the sub-network of the two communities which allows high fidelity transfer. \textit{Mean}: mean fidelity over the whole network. Each of these are calculated for each normal mode $\Omega_1-\Omega_3$ in each realization, and then the mean of means is taken over the whole ensemble. The figure is of an ensemble of 100 SBM networks.
    }
    \label{fig:topocomp}
\end{figure}

In Figure~\ref{fig:topocomp} we compare the transfer fidelities as well as the distribution of the shifted normal mode frequencies of the case of a single missing link to when a random different link has been added between the same communities. In the initial lossy case, shown in \ref{fig:topocomp}\textit{a)}, the majority of modes have their frequencies shifted significantly, while some normal modes are mostly unaffected because the coupling of the affected communities to these normal modes is small. The shifting of the normal mode frequencies causes the external oscillators to be out of resonance and leads to significantly worse transfer fidelities. When the frequency difference is small, the fidelity drop is close to linear, while at larger frequency differences the behaviour of the transfer fidelities is seemingly random. In contrast, panel \ref{fig:topocomp}\textit{b)} shows that by adding a link the majority of normal modes have their frequency restored to within a percent of the original frequencies, and subsequently the transfer fidelities are higher. 

Based on these observations, to fix the transfer network it suffices to identify the communities between which the link was lost. This should be done using less resources or operations that would be required for full network reconstruction or topology probing. We show that the shifting of the normal mode frequencies can generally be used to infer where a link has been lost. The method significantly outperforms random guessing in networks with a clear community structure and requires only probing the frequencies of the $N$ slowest normal modes \cite{nokkala2016complex,nokkala2018local}.

The amount the normal mode frequencies shift is approximately proportional to the overlap of the corresponding eigenvector with the communities between which the link was removed. No community has significant overlap with all the $N$ normal modes, so it is expected that not all normal modes are equally affected. Since it is possible to probe the network for the normal mode frequencies after the loss, and since the couplings of the nodes to the normal modes are known due to being needed in setting up the transfer, it is possible to find the most shifted normal mode frequency and the two communities in which nodes most strongly to the mode. A link can then be added between two randomly chosen nodes in these communities.

\begin{figure}
    \centering
    \includegraphics[width=0.7\linewidth]{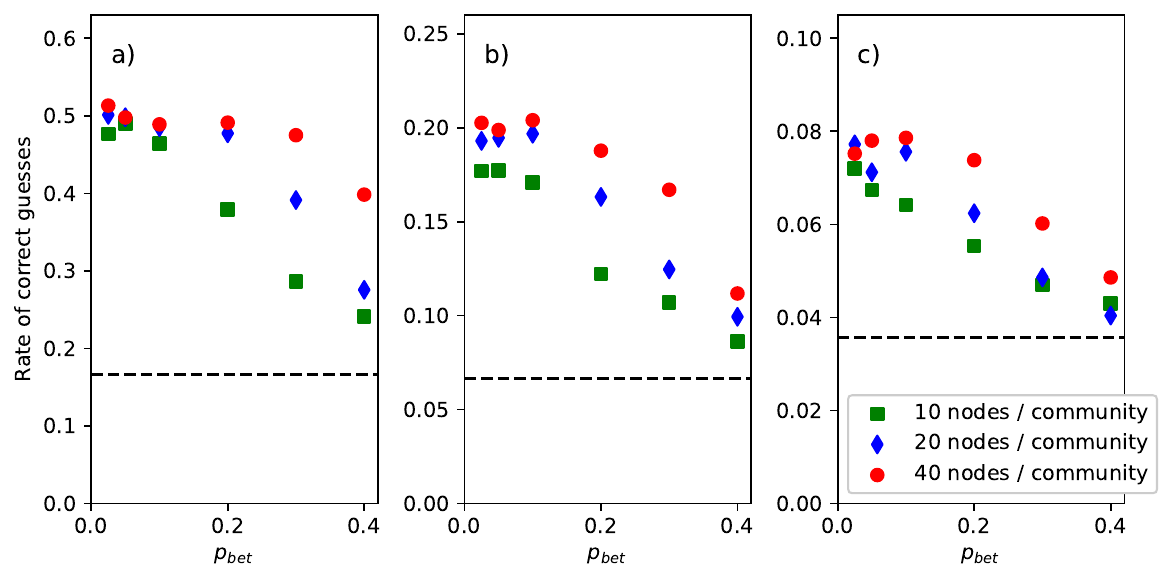}
    \caption{Identifying the communities between which a link was lost with the most shifted frequency was examined in networks of \textbf{a)}four, \textbf{b)} six and \textbf{c)} eight communities, with different community sizes and inter-community link probabilities $p_{bet}$.
    An ensemble of 5000 random networks was created for each parameter set, and a single inter-community link was removed from each. Here we show the fraction of pairs for which the communities were then correctly identified by the most shifted frequency compared to the probability of hitting the correct communities by random guessing, indicated by the horizontal line in each panel. 
    }
    \label{fig:topo_msf_comparison}
\end{figure}

In Figure~\ref{fig:topo_msf_comparison} we compare the efficiency of finding the correct communities with this method to random guessing in different sized random networks. We see that in almost all cases the method significantly outperforms random guessing. The efficiency goes down as the number of communities increases, but so does the probability of randomly guessing correctly. Similarly the efficiency decreases as $p_{bet}$ increases, i.e. when the community structure becomes less clear. The weaker community structure leads to all communities having non-insignificant overlap with all of the low normal modes, and thus removing an inter-community link generally significantly affects more than one normal mode, while simultaneously being of less overall importance due to the increased abundance of inter-community links.

While adding a link between the same communities from which the link was lost is the natural choice, it is also worth asking whether a link between any two communities would work to restore the transfer properties. This can immediately be answered in the negative by considering the coupling strengths to the $N$ slowest normal modes: inter-community links generally single out one of the normal modes as seen in Figure~\ref{fig:topofreqs}. This is the mode to which the joined communities most strongly couple to. The coupling strengths vary between communities, and a choice of a different community pair generally leads to a different normal mode being the most affected. Thus adding a link between the wrong communities will cause another normal mode's frequency to be raised by a large amount, while the most decreased frequency is only slightly increased. The result is that two of the normal modes are significantly out of tune from their expected frequencies and the transfer fidelities drop over both modes.

\section{Noise in oscillator frequencies} \label{sec:frequencynoise}

In this section we consider the case where the network has inhomogeneous frequencies, and the inhomogeneity is not known when setting up the network; transfer parameters are calculated as if the network was homogeneous. We focus on networks where a single oscillator has a different frequency from the rest, see Figure~\ref{fig:topo_info}. This simple case can be physically motivated by possible fabrication defects, and proves to have some interesting features.

\begin{figure}
    \centering
    \includegraphics[width=0.7\linewidth]{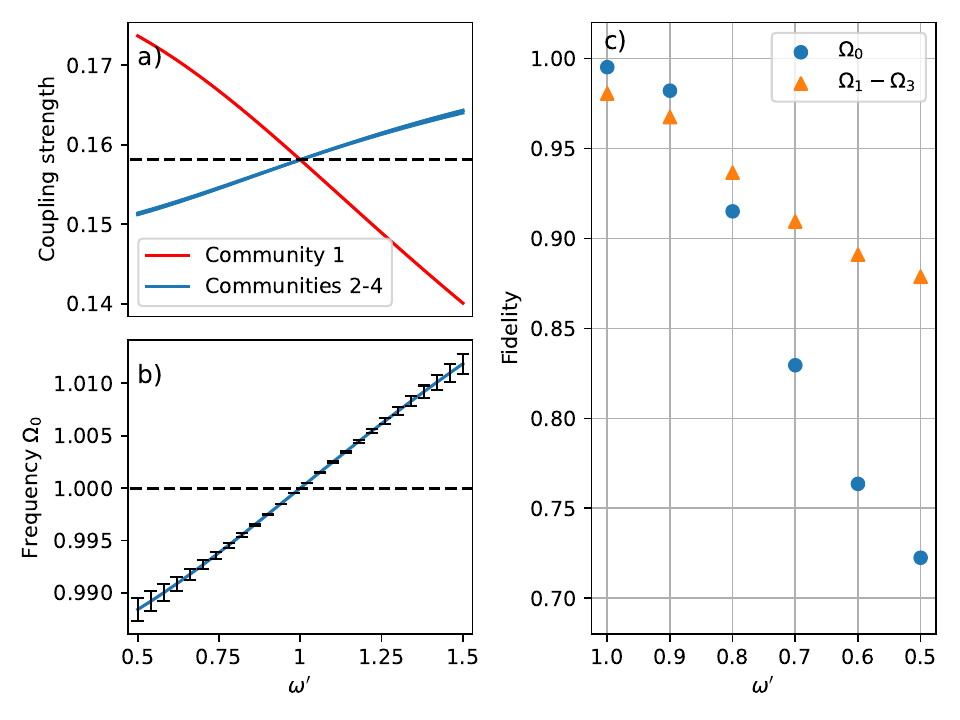}
    \caption{ One oscillator in the network has a frequency of $\omega'$ \textit{a)} Median coupling strength of nodes in each community to the slowest normal mode, with the detuned oscillator in community 1. \textit{b)} Frequency of the slowest normal mode, with error bars extending to one standard deviation from the mean. \textit{c)} Average fidelities of the best performing community over the slowest normal mode and the next few normal modes. Panels \textit{a} and \textit{b} are averaged over 1000 realizations, and panel \textit{c} is averaged over 100 realizations.
    }
    \label{fig:freq_nofix}
\end{figure}

Having a single detuned oscillator with frequency $\omega' = \omega_0 + \Delta\omega$ affects most significantly the coupling strength of all network oscillators to the slowest normal mode and its frequency $\Omega_0$, shown in Figure~\ref{fig:freq_nofix} \textit{a)} and \textit{b)}. The transfer fidelities over the slowest normal modes are shown in panel \textit{c)}, and since the effects on the network are nearly symmetrical with respect to the sign of $\Delta\omega$, for the transfer fidelities only negative values are considered. In contrast to noise in the network topology, the higher normal modes are somewhat robust towards frequency noise whereas the center-of-mass mode is more vulnerable.

Noting that the effect on coupling strength and normal mode frequency is opposite for detunings of different sign, we may devise a method for compensating for a detuned oscillator: if we know the community in which the defect is in and the frequency difference, we may tune another oscillator in the same community with an opposite detuning. Should we hit the detuned oscillator, the problem is trivially resolved. However, here we show that any oscillator in the same community suffices to mostly restore the transfer fidelities to their original levels, i.e. two oppositely detuned oscillators with frequencies $\omega' = \omega_0 + \Delta\omega$ and $\omega'' = \omega_0 - \Delta\omega$ cancel out each others effects.

\begin{figure}
    \centering
    \includegraphics[width=0.7\linewidth]{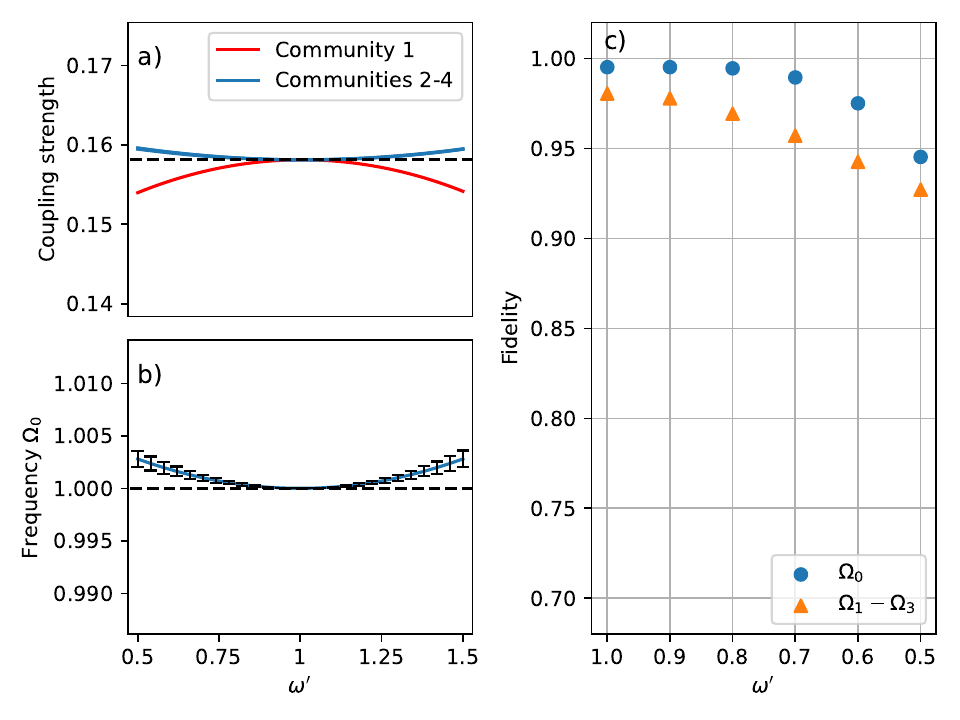}
    \caption{A single detuned oscillator with frequency $\omega' = \omega_0 - \Delta\omega$ is compensated with another oscillator tuned to $\omega'' = \omega_0 + \Delta\omega$, with both residing in community 1. \textit{a)} Median coupling strength of nodes in each community to the slowest normal mode. \textit{b)} Frequency of the slowest normal mode. \textit{c)} Average fidelities of the best performing community over the slowest normal mode and the next few normal modes. Panels \textit{a} and \textit{b} are averaged over 1000 realizations, and panel \textit{c} is averaged over 100 realizations.}
    \label{fig:freq_correctfix}
\end{figure}

Figure~\ref{fig:freq_correctfix} shows the resulting coupling strengths to the slowest normal mode as well as its frequencies are shown along with the transfer fidelities. The network properties are not perfectly restored to match the noiseless case, but rather the positive detuning has a slightly stronger effect. Thus the optimal compensatory detuning is not of equal magnitude, however the presented method is simple and not far from optimal. Additionally the transfer fidelities are mostly restored to values closer to unity even with this method.

We now turn our attention to recovering the information needed to compensate for the noise with minimal probing. To find the community in which the defective oscillator is in, we consider probing the coupling strength of the oscillators to the slowest normal mode, inspired by Figure~\ref{fig:freq_nofix}\textit{a)}. The community with the defective oscillator is clearly singled out. The figure shows the median coupling of the oscillators in each community, and thus it does not matter which oscillator we probe: the defective oscillator affects all other oscillators in its community roughly equally.

Although the coupling strength deviation is proportional to the detuning, it in general cannot be used to infer the magnitude of the detuning because to the deviations in coupling strength for a given detuning are heavily dependent on the network topology and their distribution is quite wide. However, as indicated by the error bars in Figure~\ref{fig:freq_nofix}\textit{b)}, the frequency deviation of the slowest normal mode strongly correlates with the detuned frequency. Therefore we may use the coupling strength to find out the community of the defect and the frequency of the center-of-mass mode for the defect's frequency. This then is sufficient information for the compensation method presented earlier.

Similarly to Section~\ref{sec:topologynoise} we could consider tuning an oscillator in a different community than where the original detuned oscillator was, and again it is clear by looking at Figure~\ref{fig:freq_nofix}\textit{a)} that the added detuning will not restore the network properties: a detuned oscillator affects its own community's coupling strengths the most, and by adding an opposite sign detuning to another community, the two communities would both be affected in opposite ways. On the other hand, the effect on the frequency of the slowest normal mode should be independent of the chosen community due to the constant overlap of all nodes with the corresponding eigenvector, so while the coupling strengths would be off from the value in a homogeneous network, $\Omega_0$ would be close to its initial value. The overall effect on fidelity would still be detrimental.


\section{Conclusions} \label{sec:conclusions}
Quantum state transfer has been considered in noisy networks of spins \cite{de2005perfect,rafiee2013noise,babukhin2022effect,coutinho2022robustness} and in noiseless networks of harmonic oscillators \cite{hahto2024transfer,portes2013perfect}. Here we add noise to the networks of harmonic oscillators, focusing on noise models that are inspired by the modular structure found to be beneficial to routing in \cite{hahto2024transfer}. Our results show that in general, modular networks are more robust to noise than monolithic ones. We expand on this point in Appendix~\ref{app:modular_or_chain}, showing that open or closed chains, which are more common networks studied in state transfer \cite{plenio2005high,de2005perfect,kay2016quantum}, are especially weak towards noise and do not scale well to a large number of nodes.

Specifically, we found that noise affecting the network topology mainly alters transfer fidelities over higher normal modes and the slowest normal mode is very robust towards it, while noise in the oscillator frequencies affects the slowest normal mode most and the higher modes are moderately robust. One could then carry out high fidelity transfers over the other modes in the presence of frequency noise, whereas link noise would likely hit an internal link---and even if not, one could continue to use the slowest mode. We stress that the same advantage can be expected to apply beyond the particular system considered here as long as the Hamiltonian is proportional to the Laplace matrix, such as in spin networks where links are Heisenberg interactions \cite{kay2011basics}. We also examined how the properties of the networks change due to the noise, and showed its nature can be identified from them, leading to possible ways to combat the adverse effects of noise.

In Section~\ref{sec:topologynoise} we examined how the size of the network---number of communities and their sizes---affects the rate of correctly identifying the noise in network topology. Although the results show that in networks with more communities noise is harder to identify, and thus one could argue that less communities is better for resilient routing networks, more communities means more possible normal modes over which communication can happen. Therefore a trade-off must be made in the design of the network, taking into account both the likelihood of link losses as well as the need for simultaneous transfers.

Our results could be applied to a network of spatially separated micro- or nanomechanical phonon traps, which can be simulated optically \cite{nokkala2018reconfigurable}. One may also consider a generalization to dynamic noise in temporal networks, more hardware oriented noise models, networks immersed in a heat bath, or try to search for optimal noise mitigation strategies. Finally, although only low frequency modes have substantial overlap between network nodes and can therefore be expected to properly support inter-community transfer, the usefulness of the high frequency modes to facilitate transfer between few pairs of users could be investigated.

\section*{Acknowledgements} 
M.H. acknowledges financial support from Vilho, Yrjö and Kalle Väisälä Foundation. J.N. acknowledges financial support from the Academy of Finland under project no. 348854. 

\section*{Conflict of Interest}

The authors declare no conﬂict of interest.

\bibliography{references}

\appendix

\section{On modular networks vs. open/closed chains } \label{app:modular_or_chain}

It has been shown that open or closed chains (paths or rings) of harmonic oscillators can be used for quantum state transfer and that they have other normal modes in addition to the center-of-mass mode suitable for communication \cite{plenio2005high,nicacio2016coupled}. However, we argue that a modular network is more suitable for routing due to its inherent resilience to losses of random links. 

Removing any link from a closed chain turns it into an open chain, drastically changing the network topology. Then a removal of yet another link breaks the network into disjoint components, separating the users into two groups between which communication is no longer possible. In contrast, removing a single link randomly from a modular network is likely to hit an intra-community link and thus not affect the transfer fidelities noticeably. While we showed that the removal of an inter-community link causes a non-perturbative change in the network properties, the network can be engineered in such manner that communities are connected with more than a single link between each community pair. This ensures that unless a significant amount of inter-community links are lost, the network remains connected and communication over at least the slowest normal mode stays possible.

Another point against chains comes from considering scaling to a larger network. The eigenvalues of the (unnormalized) Laplacian of an $n$-node ring network are given by $\lambda_k = 2 - 2\cos \frac{2\pi k}{n}$ for $k = 1,...,n$, which for convenience may be relabeled in ascending order \cite{chung1997spectral}. Then we immediately see that the largest eigenvalue is bound above by $\lambda_n \leq 4$, and consequently for a network of oscillators arranged in a ring the highest normal mode frequency is bound by $\Omega_n \leq 2.5$. Alongside the fact that all the eigenvalues (and normal mode frequencies) are positive, this means that as the network grows, the gaps between normal modes must get smaller. Denser normal mode spectrum means stronger coupling to normal modes adjacent to the intended one, resulting in worse transfer. Since removing a link decreases the eigenvalues \cite{porto2017eigenvalue}, in an open chain the gaps are even smaller compared to a closed chain. In contrast, for a modular network the community sizes and number of communities can be tuned to provide sufficient gaps in the spectrum.


\section{Transfer of entanglement}\label{app:entanglement}

\begin{figure}[h]
    \centering
    \includegraphics[width = 0.6\textwidth]{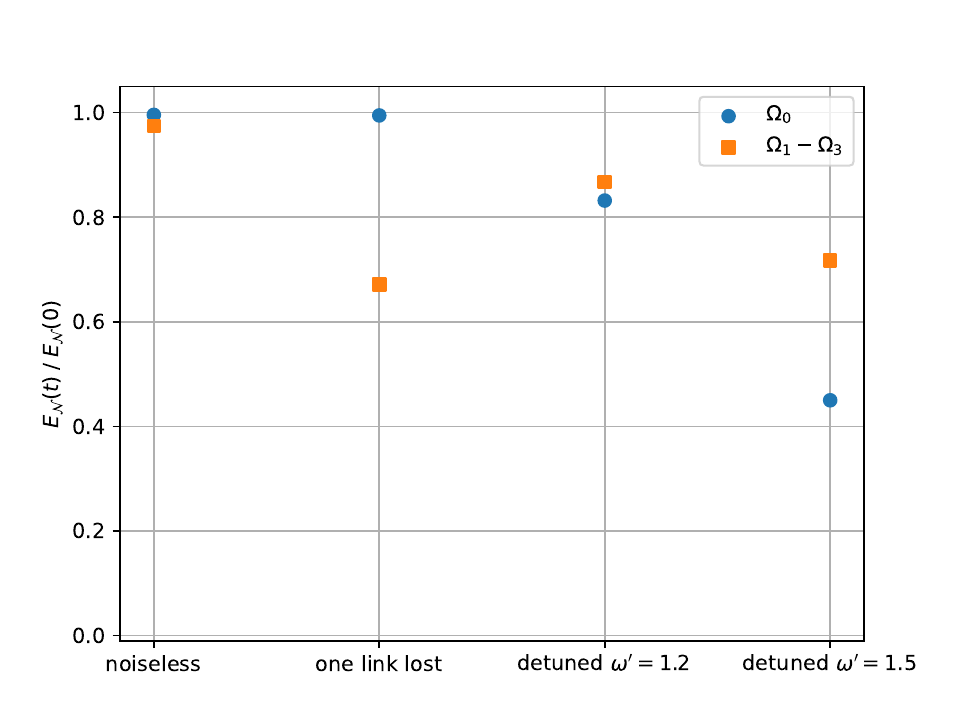}
    \caption{Fraction of transferred entanglement over the slowest normal mode $\Omega_0$ and modes $\Omega_1 - \Omega_3$, quantified by logarithmic negativity $E_\mathcal{N}$. Shown for noiseless networks, networks with a single missing inter-community link, and two cases of one detuned network oscillator with a moderate and a large detuning, $\omega' = 1.2$ and $\omega' = 1.5$ respectively. Each case is averaged over an ensemble of 100 networks.}
    \label{fig:entanglement}
\end{figure}

In \cite{hahto2024transfer} it was shown that in the ideal case where only the resonant normal mode is present, transfer fidelity becomes independent of the state as transfer time increases, justifying looking at only transfer of one possible initial state. Here we confirm that the results obtained in the main text for noisy networks and a squeezed state also hold for the transfer of entanglement by considering a two mode squeezed vacuum as the initial state. One of the initially entangled modes is the sender oscillator coupled to the network, while the other is an auxiliary oscillator disconnected from the rest. Transfer of entanglement is then quantified by the logarithmic negativity $E_\mathcal{N}$ (calculated using the QuTiP Python library \cite{johansson2012qutip}) between the receiving oscillator and the auxiliary oscillator at time $t_{ideal}$. Figure \ref{fig:entanglement} shows the behaviour of $E_\mathcal{N}$ for the different types of noise, and we see a familiar pattern: for noise in network topology transfer over $\Omega_0$ is mostly unaffected while over $\Omega_1 - \Omega_3$ the transfer quality declines. For noise in oscillator frequencies the transfer efficiency drops sharply over $\Omega_0$ as the detuning increases, while the drop over $\Omega_1 - \Omega_3$ is slower.

\end{document}